# Earth's surface fluid variations and deformations from GPS and GRACE in global warming

Shuanggen Jin[1,*], Liangjing Zhang[1,2], Guiping Feng[1,2]

[1]Shanghai Astronomical Observatory, Chinese Academy of Sciences, Shanghai 200030, China
[2]Graduate University of the Chinese Academy of Sciences, Beijing 100049, China
*Corresponding author, e-mail: sgjin@shao.ac.cn; shuanggen.jin@gmail.com

*Abstract*—Global warming is affecting our Earth's environment. For example, sea level is rising with thermal expansion of water and fresh water input from the melting of continental ice sheets due to human-induced global warming. However, observing and modeling Earth's surface change has larger uncertainties in the changing rate and the scale and distribution of impacts due to the lack of direct measurements. Nowadays, the Earth observation from space provides a unique opportunity to monitor surface mass transfer and deformations related to climate change, particularly the global positioning system (GPS) and the Gravity Recovery and Climate Experiment (GRACE) with capability of estimating global land and ocean water mass. In this paper, the Earth's surface fluid variations and deformations are derived and analyzed from global GPS and GRACE measurements. The fluids loading deformation and its interaction with Earth system, e.g., Earth Rotation, are further presented and discussed.

*Key words*-Surface fluids; deformation; GPS; GRACE

## I. INTRODUCTION

Due to global warming, the Earth's surface fluid mass was redistributed, such as the glacier melting, land water runoff and sea level rise, which will result in changes in Earth's surface loading and deformation as well as tectonic activities. However, observing and modeling Earth's surface mass change has larger uncertainties in the changing rate and the scale and distribution of impacts due to the lack of direct measurements. Since its launch in March 2002, the Gravity Recovery and Climate Experiment (GRACE) mission has been widely used to estimate the Earth's time-variable gravity field by accurately determining the relative position and changing rate of a twin of Low Earth Orbit (LEO) satellites [1]. Gravity changes reflect the total mass redistribution and motion within Earth system, including the land and ocean water and atmosphere pressure, so the GRACE provide a unique opportunity to directly measure the transfer and motion of the Earth's surface fluid mass [2]. The Global Positioning System (GPS) as a real-time, high precision and global covering technique has provided an unprecedented high accuracy and tremendous contribution to positioning and scientific questions related to precise positioning on the Earth's surface, since it became fully operational in 1993. It monitors the millimeter-level motion and deformation of the Earth's crust due to the Earth's surface mass redistribution and motion [3, 4]. Therefore, the GPS and GRACE both detect surface mass transfer, deformations and trends related to climate change as well as validate each other.

In addition, the mass redistribution and movement in the atmosphere, oceans and hydrosphere in the current climate change also lead to the Earth's rotational changes at different time scales. However, quantitative assessment of the Earth's surface fluid mass contributions to polar motion and length-of-day (LOD) remains unclear due mainly to the lack of global direct observations, particularly the terrestrial water and ocean bottom pressure data [2]. In this paper, the Earth's surface fluid variations and deformations are derived and analyzed from global GPS and GRACE observations. The fluids loading deformation and its interaction with Earth system, e.g., Earth Rotation, are further presented and discussed.

.

## II. DATA AND METHDS

Continous IGS GPS observations can monitor crustal and loading deformation due to surface fluid mass redistribution and tectonic activities. Currently global IGS network has more than 300 sites with observations from 1994.0 to present. Here the time series of IGS coordinate are used from the International Terrestrial Reference Frame 2005 (ITRF2005). These solutions are provided through combining each IGS Analysis Center (AC) solution. The Solid Earth tides, ocean tide and pole tide have been modeled in the original data processing. The GPS sites with larger noisy and big gaps were removed and finally 160 sites were chosen.

The time-varying GRACE gravity fields can be used to estimate high-quality terrestrial water and ocean mass change. For example, monthly GRACE gravity changes over oceanic regions can be transformed to ocean mass redistribution or ocean

Supported by the National Natural Science Foundation of China and the program of Chinese Academy of Sciences

bottom pressure (OBP) at colatitude $\varphi$, longitude $\lambda$ as described by *Wahr et al.* [5]:

$$Mass(\phi,\lambda) = \frac{ag\rho_E}{3} \sum_{l=2}^{L} \sum_{m=0}^{l} \frac{2l+1}{1+k_l} W_l \tilde{P}_{lm}(\cos\phi) \quad (1)$$
$$[C_{lm}\cos(m\lambda) + S_{lm}\sin(m\lambda)]$$

where $a$ is the earth's semimajor axis radius, $g$ is the mean gravitational acceleration, $\rho_E$ is the mean density of the Earth, $k_l$ is the loading Love numbers of degree $l$, $W_l$ is the Gaussian averaging function, $\tilde{P}_{lm}$ is the fully normalized associated Legendre functions of degree $l$ and order $m$, and $C_{lm}$ and $S_{lm}$ are the component of the GRACE spherical harmonic coefficients of degree $l$ and order $m$. The high-precision GRACE gravity field solutions (Release-04) are used from the University of Texas Center for Space Research (UTCSR) from August 2002 to December 2008. In order to be consistent with reference frame for GPS data, the degree *1* term was added along with GRACE GSM coefficients for degrees 2 and higher. In addition, since the GRACE is not sensitive to degree 2, the $C_{20}$ coefficient was replaced from Satellite Laser Ranging data [6]. The highly precise $1°\times1°$ monthly OBP and Terrestrial Water Storage (TWS) time series can be estimated from the time dependent component of the GRACE spherical harmonic coefficients [7]. The coefficients suffer from a correlated error that causes 'stripes' in the maps, such as TWS in Jan. 2003 (Figure 1)

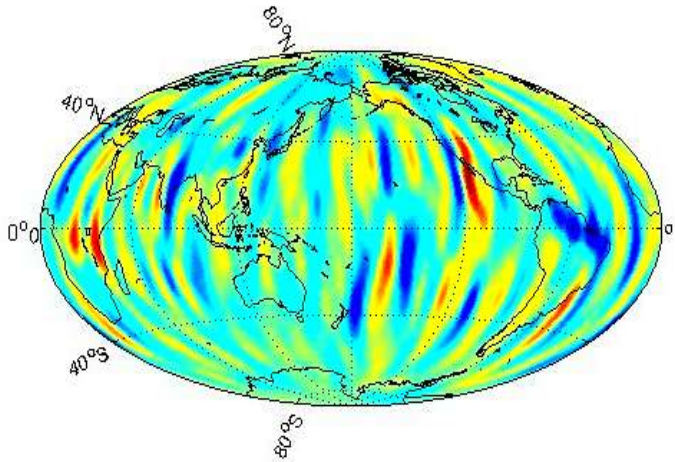

Figure 1. Terrestrial Water Storage (TWS) distribution without de-striping in January 2003.

Here the monthly grid OBP and water storage data are filtered with a 500-km Gaussian smooth from August 2002 until January 2009, except for missing data in June 2003 and January 2004. For example, Figure 2 denotes the distribution of terrestrial water storage (TWS) minus the mean (equivalent water thickness change in millimeter) in January 2003 and Figure 3 shows the distribution of OBP minus the mean (equivalent water thickness residual in centimeter) in August 2002.

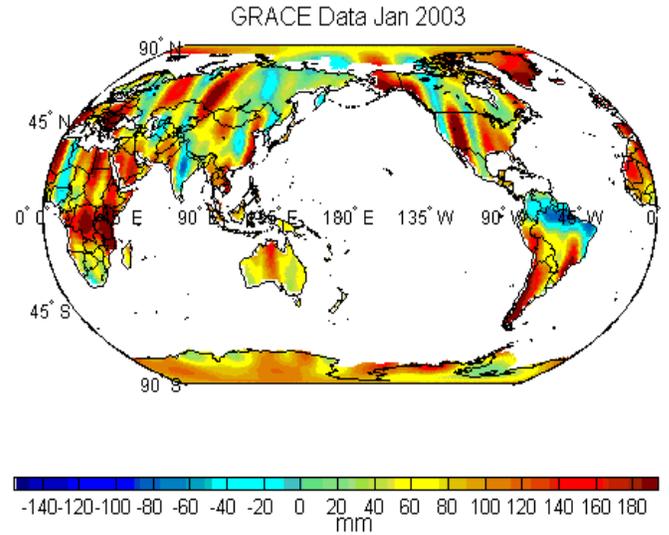

Figure 2. The distribution of terrestrial water storage (TWS) minus the mean (equivalent water thickness change in millimeter) in January 2003.

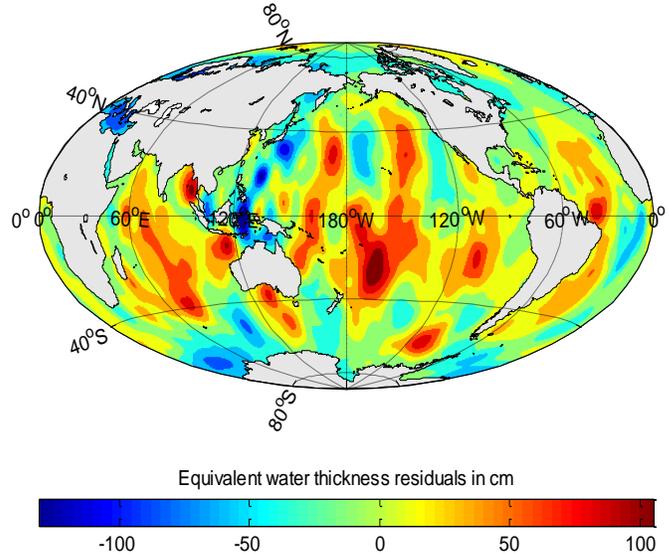

Figure 3. The distribution of OBP minus the mean (equivalent water thickness change in centimeter) in August 2002.

### III. RESULTS AND DISCUSSION

The Earth's surface fluid mass variations and trends are estimated and analyzed from monthly GRACE solutions, which reflect mass loading and deformations due to surface fluid mass transfer in climate change, such as ice melting and sea level rise. For example, Figure 4 shows the similar trend distributions of ocean bottom pressure as equivalent water thickness variation in cm/yr from the GRACE and ocean model ECCO. These variation trends are due mainly to the input of land fresh water and glacial melting.

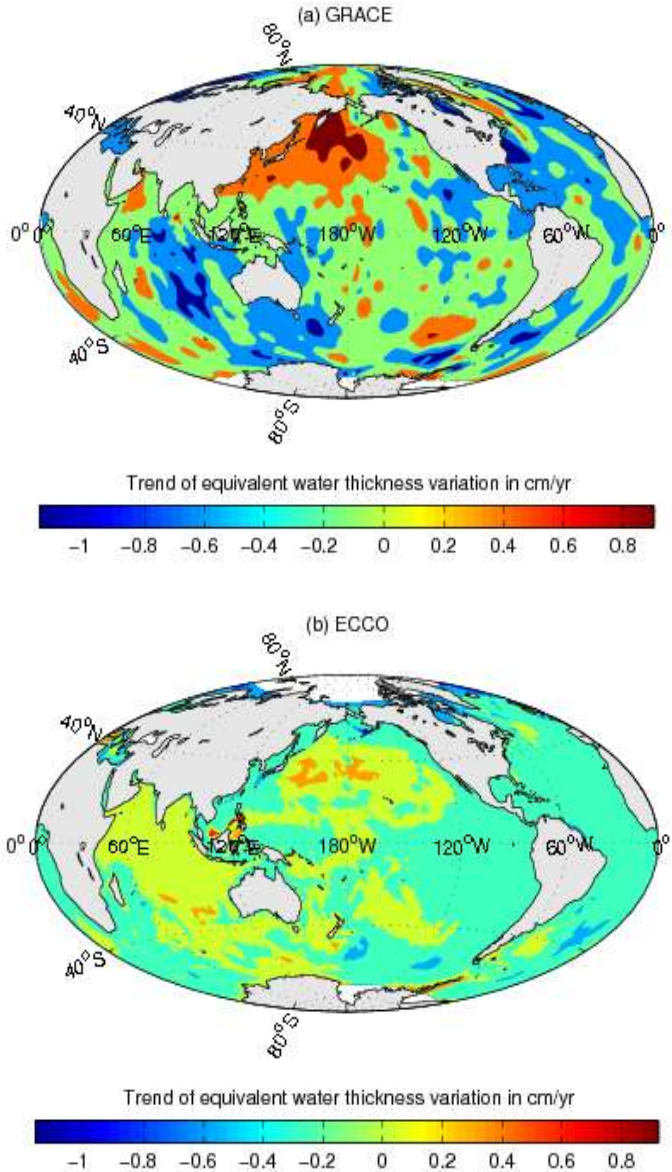

Figure 4. The trend distributions of ocean bottom pressure as equivalent water thickness variation in cm/yr from GRACE and ECCO model.

Meanwhile the loading deformations can be transferred into displacement, which could validate the GPS direct measurements. An annual signal and semi-annual signal have been fitted by weighted least squares for each site coordinate residual time series by GPS and GRACE. The annual amplitude $A_1$ and phase $\phi_1$ are defined as

$$a + bt + A_1 \sin(\omega_1(t - t_o) + \phi_1) + A_2 \sin(\omega_2(t - t_o) + \phi_2) \quad (2)$$

where $t_o$ is 2003.0 and $\omega_1$ is the angular frequency of 1 cycle/yr and $\omega_2$ is the angular frequency of 0.5 cycle/yr. Figure 5 shows the amplitude and phase of the annual signal of the GPS and GRACE vertical coordinate time series. The annual signals from most GPS sites agree with GRACE, e.g., located in Asia, Africa and South America. However, the discrepancies are also found at some sites, especially those located at islands and coastal regions. We also compare the displacements from the geophysical data, while it is worse than GRACE measurements.

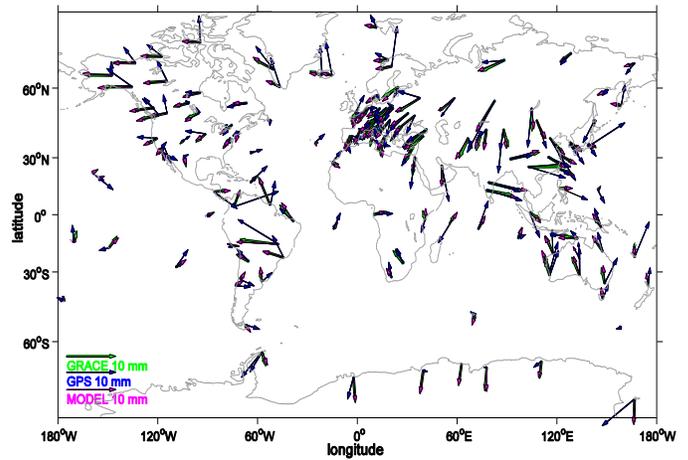

Figure 5. Comparison of annual signals from GPS, GRACE and geophysical models. Annual amplitudes and phases of the vertical components of GPS and GRACE surface displacements. The length of the arrow represents the amplitude of the signals, and the phase is counted counterclockwise from the east. Green arrow is the GRACE annual signals, blue one is GPS estimate and purple one shows the geophysical model estimate.

Furthermore, statistical parameters are computed for a more quantitative analysis and comparison with the RMS of the GPS heights minus GRACE predicted heights. After removing the GRACE predictions the RMS of GPS height residuals are reduced at most sites. However, it still has some inconsistency. On the one hand, the used GPS data are the combination of height time series derived from various IGS analysis centers with different software and processing techniques, which will result in errors in GPS coordinate time series. In addition, the GPS mapping function and phase center variation model also have larger uncertainties, which can lead to the spurious annual signals in the GPS solutions [8, 9]. On the other hand, GRACE has a low spatial resolution, which cannot map small basin mass variations. Moreover, there are lots of larger uncertainties in GRACE solutions, particularly the filter methods. It needs the future high resolution gravimetric satellite missions.

Additionally, the redistribution of Earth's surface mass will result in variations of polar motion and length-of-day (LOD), including atmospheric angular momentum (AAM), oceanic angular momentum (OAM), and hydrological angular momentum (HAM). While these excitation angular momentums are normally from geophysical models, such as the Estimating Circulation and Climate of the Ocean (ECCO) model. Unfortunately, these excitation results rely on geophysical models with relatively few observational data input [2]. The direct GRACE observations provide a new opportunity to investigate and understand surface fluid mass contributions to polar motion and length-of-day (LOD). For example, Figure 6 shows Monthly Earth's surface mass excitation time series from non-atmospheric wind and ocean currents geodetic observation residuals LOD (blue), models' estimates AOH (black), SLR estimates (green) and GRACE estimates (cyan) as well as GRACE/SLR combined solutions (red). The excitations from GRACE and SLR observations are better than models' estimates in explaining the geodetic residuals of non-wind/currents excitations at the annual period, while the excitation from combined GRACE and SLR is much improved in explaining geodetic residual LOD [10].

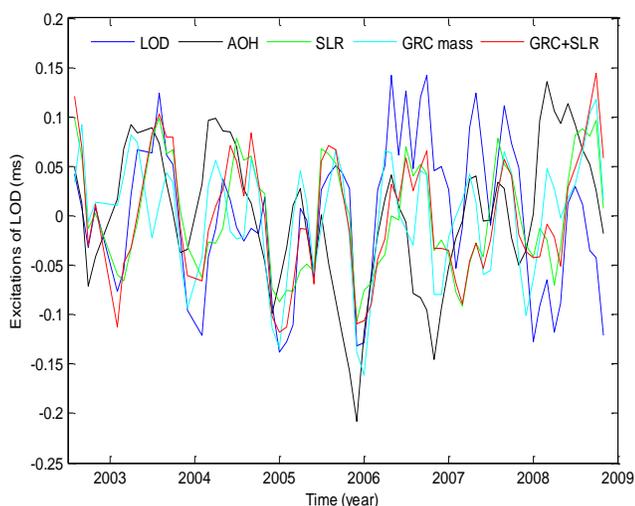

Figure 6. Monthly Earth's surface mass excitation time series from non-wind/currents geodetic observation residuals LOD (blue line), models' estimates AOH (AAMp+OAMp+HAM) (black line), SLR estimates (green line), GRC mass estimates (cyan line) and combined SLR and GRC estimates (red line).

## IV. CONCLUSION

The Gravity Recovery and Climate Experiment (GRACE) mission provides the time-varying gravity filed, which directly estimate the mass variations within the Earth system and loading deformation. The Earth's surface fluid mass variations and trend are analyzed and compared from monthly GRACE solutions and geophysical models, showing similar variability and trend. For example, the ocean bottom pressure increase was due mainly to the input of land fresh water and glacial melting in current global warming. The surface fluid mass redistribution from GRACE also results in loading deformations, which are further analyzed and compared by GPS measurements at the global scale with 160 IGS sites. At most sites located in Asia, Africa and South America, good agreements are found, indicating that the Earth's surface deformation caused by geophysical mass loading may explain the seasonal variations of the GPS coordinate time series. However, some discrepancies are also shown at islands and coastal sites. It needs to further investigate with future high resolution observations. In addition, the redistribution of Earth's surface mass results in variations of polar motion and length-of-day (LOD). Analysis results show that the excitations from GRACE observations are better than models' estimates in explaining the geodetic residuals of non-wind/currents excitations of polar motion and length-of-day (LOD) variation at the annual scale, but still needs to be improved at high frequency variations.

## ACKNOWLEDGMENT

The authors would also like to thank GRACE data and IGS who provide GPS data as well as other helps.